%% file: ANA-TOPQ-2021-24-PAPER.tex
\documentclass[PAPER, atlasdraft=false, texlive=2023, USenglish, coverpage=false, cernpreprint=true, orcidlogo, texmf, orcidlogo]{atlasdoc}
 
\usepackage[subfigure=true,block=space]{atlaspackage}
\usepackage{atlasbiblatex}
\usepackage{braket}
\usepackage{tikz-feynman}
\usepackage{dsfont}
\usepackage{xcolor}
 
\usepackage{atlasphysics}

\newcommand*{\ttH}{\ensuremath{t\bar{t}H}\xspace}
\newcommand*{\ttV}{\ensuremath{t\bar{t}V}\xspace}

\newcommand*{\ttll}{\ensuremath{t\bar{t}\ell\ell}\xspace}
\newcommand*{\ptmiss}{\ensuremath{\vec{p}_{\text{T}}^{\text{miss}}}\xspace}
 
\newcommand*{\cosphi}{\ensuremath{\cos\varphi}\xspace}
 
\graphicspath{{logos/}{figures/}}
 
\usepackage{ANA-TOPQ-2021-24-PAPER-defs}
\usepackage{amsmath, bm}

\AtlasTitle{Observation of quantum entanglement with top quarks at the ATLAS detector}

\AtlasAbstract{
Entanglement is a key feature of quantum mechanics~\cite{Einstein:1935rr,Schrodinger1935,Bell:1964kc}, with applications in fields such as metrology, cryptography, quantum information, and quantum computation~\cite{Bennett2000,Nielsen2000,Marciniak2022,Horodecki:2009zz,Casini:2022rlv}.
It has been observed in a wide variety of systems and length scales, ranging from the microscopic~\cite{Aspect1982,Hagley1997,Steffen2006,Pfaff2013,Belle:2007ocp} to the macroscopic~\cite{Julsgaard2001,Lee2011,Ockeloen2018}.
However, entanglement remains largely unexplored at the highest accessible energy scales.
Here we report the highest-energy observation of entanglement, in top--antitop quark events produced at the Large Hadron Collider, using a proton--proton collision dataset with a center-of-mass energy of $\sqrt{s}=13$~\TeV\ and an integrated luminosity of 140~fb$^{-1}$ recorded with the ATLAS experiment.
Spin entanglement is detected from the measurement of a single observable $D$, inferred from the angle between the charged leptons in their parent top- and antitop-quark rest frames.
The observable is measured in a narrow interval around the top--antitop quark production threshold, where the entanglement detection is expected to be significant.
It is reported in a fiducial phase space defined with stable particles to minimize the uncertainties that stem from limitations of the Monte Carlo event generators and the parton shower model in modeling top-quark pair production.
The entanglement marker is measured to be $D=-0.537 \pm 0.002~\text{(stat.)} \pm 0.019~\text{(syst.)}$ for $340 < \mttbar < 380 $~\GeV.
The observed result is more than five standard deviations from a scenario without entanglement and constitutes the first observation of entanglement in a pair of quarks and the highest-energy observation of entanglement so far.
}

\AtlasRefCode{TOPQ-2021-24}
 
\PreprintIdNumber{CERN-EP-2023-230}
 
\AtlasDate{\today}

\AtlasJournalRef{Nature 633 (2024) 542}
\AtlasDOI{10.1038/s41586-024-07824-z}

\hypersetup{pdftitle={ATLAS document},pdfauthor={The ATLAS Collaboration}}
 
\begin{document}
 
\maketitle
 
\tableofcontents
 
\clearpage

\section{Main}
\label{sec:intro}

 
Particle colliders, such as the Large Hadron Collider (LHC) at CERN, probe fundamental particles and their interactions at the highest energies accessible in a laboratory, exceeded only by astrophysical sources.
Beyond the fundamental interest of exploring quantum entanglement in a novel setting, this observation demonstrates the potential of using high-energy colliders, such as the LHC, as tools for testing our fundamental understanding of quantum mechanics.
Hadron colliders offer a truly relativistic environment and provide a rich menu of fundamental interactions, rarely considered for experiments in quantum information.
Relativistic effects are expected to play a critical role in quantum information~\cite{Peres:2002wx} and the measurement described here illustrates the potential for novel approaches to explore these effects and other foundational problems in quantum mechanics using colliders.

Recently, the heaviest fundamental particle known to exist, the top quark, was proposed as a new laboratory to study quantum entanglement and quantum information~\cite{Afik:2020onf,Afik:2022kwm}.
In this Article, the spin correlation between the top quark and antitop quark is used to probe the effects of quantum entanglement, in proton--proton ($pp$) collision events recorded with the ATLAS detector at a center-of-mass energy of 13~\TeV.
Entanglement is observed with a significance of more than
five standard deviations for the first time in pairs of quarks.
 
If two particles are entangled, the quantum state of one particle cannot be described independently of the other.
The simplest example of an entangled system involves a pair of quantum bits (qubits); pieces of quantum information about two particles in the same quantum state which exist in superposition.
The spin quantum number of a fundamental fermion, a particle whose spin can take values of $\pm 1/2$, is one of the simplest and most fundamental examples of a qubit.
Among the fundamental fermions of the Standard Model (SM) of particle physics, the top quark is uniquely suited for high-energy spin measurements because of its unique properties: its immense mass gives it a lifetime (${\sim}10^{-25}$\,s) notably shorter than the timescale needed for a quark's quantum numbers to be shrouded by hadronization (${\sim}10^{-24}$\,s) and spin decorrelation (${\sim}10^{-21}$\,s) effects~\cite{ParticleDataGroup:2022pth}.
As a result, its spin information is transferred to its decay products.
This unique feature provides an opportunity to study a pseudo-bare quark, free of the color-confinement properties of the strong force that shroud other quarks.
 
Quarks are most commonly produced in hadron collider experiments as matter--antimatter pairs. A pair of top--antitop quarks (\ttbar) is a two-qubit system whose spin quantum state is described by the spin density matrix $\rho$:
\begin{equation*}\label{eq:GeneralBipartiteStateRotations}
\rho=\frac{1}{4} \left[ I_4+\sum_{i} \left( B^{+}_{i}\sigma^i\otimes I_2+B^{-}_{i}I_2\otimes\sigma^i \right) + \sum_{i,j}C_{ij}\sigma^{i}\otimes\sigma^{j} \right] \text{.}
\end{equation*}
 
The first term in the linear sum is a normalization constant, where $I_n$ is the $n\times n$ identity matrix. The second term describes the intrinsic polarization of the top and the antitop quarks, where $\sigma^i$ are the corresponding Pauli matrices and the real numbers $B^{\pm}_{i}$ characterize the spin polarization of each particle. The third term describes the spin correlation between the particles, encoded by the spin correlation matrix $C_{ij}$. In all expressions, an orthogonal coordinate system is represented by the indices $i,j=1,2,3$.
 
At hadron colliders, \ttbar pairs are produced mainly via the strong interaction and thus have no intrinsic polarization (i.e.\ $B^{\pm}_{i} \simeq 0)$ because of parity conservation and time invariance in quantum chromodynamics (QCD)~\cite{Mahlon:1995zn}. However, their spins are expected to be correlated and this correlation has already been observed by both the ATLAS and CMS experiments at the LHC~\cite{TOPQ-2011-11,TOPQ-2014-07,TOPQ-2016-10,CMS-TOP-13-003,CMS-TOP-18-006}.
Entanglement in top-quark pairs can be observed via an increase in the strength of their spin correlations.

Due to their short lifetime, top quarks cannot be detected directly in experiments.
In the SM, they decay almost exclusively into a bottom quark and a $W$ boson, and the $W$ boson subsequently decays into either a pair of lighter quarks or a charged lepton and a neutrino.
In this measurement, only $W$ bosons decaying into leptons are considered since charged leptons, especially electrons and muons, are readily detected with high precision at collider experiments.
To a good approximation, the degree to which the leptons carry the spin information of their parent top quarks is 100\% due to the maximally parity-violating nature of the electro-weak charged current. The angular direction of each of these leptons is correlated with the direction of the spin of their parent top quark or antitop quark in such a way that the normalized differential cross-section ($\sigma$) of the process may be written as~\cite{Bernreuther:1997gs}:
\begin{equation*}\label{eq:LeptonicCrossSection}
\frac{1}{\sigma}\frac{\mathrm{d}\sigma}{\mathrm{d}\Omega_{+}\mathrm{d}\Omega_{-}}=\frac{1+\mathbf{B}^{+}\cdot\hat{\mathbf{q}}_{+}-\mathbf{B}^{-}\cdot\hat{\mathbf{q}}_{-}
-\hat{\mathbf{q}}_{+}\cdot \mathbf{C} \cdot\hat{\mathbf{q}}_{-}}{(4\pi)^2}~,
\end{equation*}
where $\hat{\mathbf{q}}_{+}$ is the antilepton direction in the rest frame of its parent top quark and $(\hat{\mathbf{q}}_{-})$ is the lepton direction in the rest frame of its parent antitop quark; and $\Omega_{+}$ is the solid angle associated with the antilepton and $(\Omega_{-})$ is the solid angle associated with the lepton. The vectors $\mathbf{B}^{\pm}$  determine the top-quark and antitop-quark polarizations, whereas the matrix $\mathbf{C}$ contains their spin correlations.
These terms are the same as those that appear in the general form for $\rho$.
Since the information about the polarizations and spin correlations of the short-lived top quarks is transferred to the decay leptons, their values can be extracted from a measurement of angular observables associated with these leptons, allowing us to reconstruct the \ttbar spin quantum state.

The experiments at the LHC ring, such as ATLAS, are the only ones currently taking data which is able to produce and study the properties of the top quark.
At the LHC, \ttbar~pairs are produced mainly via gluon--gluon fusion. 
When they are produced close to their production threshold, i.e.\ when their invariant mass \mttbar~is close to twice the mass of the top quark ($\mttbar \sim 2\cdot m_t \sim 350$~\GeV), approximately 80\% of the production cross-section of \ttbar~pairs arises from a spin-singlet state~\cite{Kiyo:2008bv,Kuhn:1992qw,Petrelli:1997ge}, which is maximally entangled.
After averaging over all possible top-quark directions, entanglement only survives close to threshold because of the rotational invariance of the spin-singlet. 
This invariance implies that the trace (the sum of all of the diagonal elements) of the correlation matrix $\mathbf{C}$, where each diagonal element corresponds to the spin correlation in a particular direction, is a good entanglement witness. It is an observable that can signal the presence of entanglement, with $\textrm{tr}[\mathbf{C}]+1<0$ as a sufficient condition for entanglement~\cite{Afik:2020onf}.
It can be understood as a violation of a Cauchy--Schwarz inequality, a notable entanglement criterion in fields such as quantum optics, condensed matter or analogue gravity~\cite{PhysRevD.9.853,PhysRevA.73.012110, Wolk2014,deNova:2012hm}

It is more convenient to define an entanglement marker by using $D=\textrm{tr}[\mathbf{C}]/3$~\cite{Afik:2020onf}, which can be experimentally measured as:
\begin{equation*}
D =~-3\cdot \langle \cosphi \rangle~,
\label{eq:EntPrediction}
\end{equation*}
where $\langle \cosphi \rangle$ is the average value of the cosine of the angle (dot product) between the charged-lepton directions after they have been Lorentz boosted into the \ttbar\ rest frame and then their parent top-quark and antitop-quark's rest frames, which can be measured experimentally in an ensemble data set.
The existence of an entangled state is demonstrated if the measurement satisfies $D < -1/3$, derived from the Peres--Horodecki criterion~\cite{Peres:1996dw,Horodecki:1997vt}, and is independent from the order of the calculation.
It should be noted that the CMS collaboration has already measured $D = -0.237 \pm 0.011$~\cite{CMS-TOP-18-006} inclusively, showing no signal of entanglement.
 
The SM is a quantum theory and entanglement is implicitly present in its predictions. Nevertheless, a demonstration of spin entanglement in \ttbar\ pairs is challenging due to the inability to control the internal degrees of freedom in the initial state~\cite{Afik:2022kwm}. Currently, entanglement can only be detected with the help of a dedicated analysis in a restricted phase space like the one presented here.
 
\subsection{The ATLAS detector and event samples}
\label{sec:detector}
 
The ATLAS experiment~\cite{PERF-2007-01,ATLAS-TDR-19,PIX-2018-001} at the LHC is a multipurpose particle detector with a forward--backward symmetric cylindrical geometry and a solid-angle coverage of almost $4\pi$.
It is used to record particles produced in LHC collisions through a combination of particle position and energy measurements. The coordinate system is defined Methods~\ref{sec:atlas}.
It consists of an inner-tracking detector surrounded by a thin superconducting solenoid providing a $2$~T axial magnetic field, electromagnetic and hadronic calorimeters, and a muon spectrometer.
The muon spectrometer surrounds the calorimeters and is based on three large superconducting air-core toroidal magnets with eight coils each providing a field integral of between 2.0 and 6.0 T m across the detector.
An extensive software suite~\cite{ATL-SOFT-PUB-2021-001} is used in data simulation, in the reconstruction
and analysis of real and simulated data, in detector operations, and in the trigger and data acquisition
systems of the experiment.
The complete data set of $pp$ collision events at a center-of-mass energy of \sqsthirt collected with the ATLAS experiment during 2015--2018 is used, corresponding to an integrated luminosity of 140~\ifb.
This analysis focuses on the data sample recorded using single-electron or single-muon triggers~\cite{TRIG-2016-01}.
 
A unique feature of particle physics is that very precise simulations of the SM can be realized through the use of Monte Carlo (MC) event generators. These simulations replicate real collisions and their resultant particles on an event-by-event basis and these events can be passed through sophisticated simulations of the ATLAS detector to produce simulated data. Comparing these simulated events to those actually recorded by the detector is one way to test the predictions of the SM. Another is to use the simulated data to model how the ATLAS detector responds to a particular physics process, such as the pair production of top quarks, and to use this to create corrections to undo the effect of the detector response on real data and then to compare this corrected data to theoretical predictions. This measurement uses the latter strategy.
 
Three distinct types of real and simulated data are used, each with associated physics objects. Detector level refers to real data before it has been corrected for detector effects and simulated data after they have been passed through simulation of the ATLAS detector. Parton level refers to simulated MC events where the particles arise from the fundamental interaction being simulated, such as quarks and bosons, or to real collision data that has been corrected to this level. Particle level refers to simulated data with physics objects that are built only from the stable particles that remain after the decay of the particles that exist at parton level, i.e. particles that live long enough to interact with the detector, or to real data that has been corrected to this level. This measurement relies on the selection and reconstruction of muons, electrons, quarks and gluons as hadronic jets, neutrinos as missing transverse momentum (\ptmiss), $W$ bosons and top quarks. These objects are each reconstructed at detector level, particle level, and parton level. Details of how these objects are reconstructed in ATLAS and in MC simulations are provided in Methods~\ref{sec:atlas}.

MC event simulations are used to model the \ttbar signal and the expected SM background processes. The production of \ttbar\ events was modeled using the \POWHEGBOX[v2]~heavy-quark (hvq)~\cite{Frixione:2007nw,Nason:2004rx,Frixione:2007vw,Alioli:2010xd} generator at next-to-leading order (NLO) precision in QCD and the events were interfaced to either \PYTHIA[8.230]~\cite{Sjostrand:2014zea} or \HERWIG[7.2.1]~\cite{Bahr:2008pv,Bellm:2019zci} to model the parton shower and hadronization. The decays of the top quarks, including their spin correlations, were modeled at leading-order (LO) precision in QCD. An additional sample that generates \ttbar\ events at full NLO accuracy in production and decay was generated using the \POWHEGBOXRES~\cite{Jezo:2015aia,Jezo:2016ujg} ($bb4\ell$) generator, interfaced to \PYTHIA. Further details of the setup and tuning of these generators are provided in Methods~\ref{sec:MonteCarlo}. An important difference between \PYTHIA and \HERWIG is that the former uses a \pt-ordered shower, while the latter uses an angular-ordered shower (see Methods~\ref{appendix:partonshower}). Another important consideration is that full information on the spin density matrix is not passed to the parton shower programs and therefore is not fully preserved during the shower.

The SM background processes which contribute to the analysis are the production of a single top quark with a $W$ boson ($tW$), pair production of top quarks with an additional boson $t\bar{t}+X$ ($X=H,W,Z$), and the production of dileptonic events from either one or two massive gauge bosons ($W,Z$ bosons).
The generators for the hard-scatter processes and the showering are listed in Methods~\ref{sec:MonteCarlo}. 
The procedure for identifying and reconstructing detector-level objects are the same for data and MC events.

\subsection{Analysis procedure}
\label{sec:analysis}
 
Only events taken during stable-beam conditions, and for which all relevant components of the detector were operational, are considered.
To be selected, events must have exactly one electron and one muon with opposite-sign electric charges.
A minimum of two jets is required, and at least one of them must be identified to originate from a $b$-hadron ($b$-tagged).
 
The background contribution of events with reconstructed objects that are misidentified as leptons, referred to as the \enquote{fake-lepton} background, is estimated using a combination of MC prediction and correction based on data.
This data-driven correction is obtained from a control region dominated by fake leptons.
It is defined by using the same selection criteria as above, except that the two leptons must have same-sign electric charges. The difference between the numbers of observed events and predicted events in this region is taken as a scale factor and applied to predicted fake-lepton events in the signal region.

Events that pass the event selection are separated into three analysis regions, based on either the detector-level, particle-level, or parton-level \mttbar, depending on the region.
The signal region is constructed to be dominated by events that are as close to the production threshold as the resolution of the reconstruction method will allow, as this is the region where the entanglement of the top quarks is expected to be maximized.
 
The optimal mass window for the signal region was determined to be $340 < \mttbar < 380$~\GeV.
Two additional validation regions are defined in order to validate the method used for the measurement.
Firstly, a region is defined close to the limit where entanglement is not expected to be observable, and also with sizeable dilution from misreconstructed events from non-entangled regions, by requiring $380 < \mttbar < 500$~\GeV.
Secondly, a region in which no signal of entanglement is expected is defined with $\mttbar > 500$~\GeV.
Each of the regions has a \ttbar-event purity of more than 90\%.
The dominant sources of background processes arise from $tW$ and fake-lepton, accounts for 56\% and 27\% of the background in the signal region, respectively.
The remaining 17\% of background events arise from $t\bar{t}+X$ and the production of dileptonic events from either one or two massive gauge bosons.
The distribution of \cosphi in the signal region and the detector-level $D_{\mathrm{detector}}$ value, built from the \cosphi at the reconstructed detector-level and after background subtraction, are shown in the left and right panels of Figure~\ref{fig:cosphi}, respectively. 
 
In order to compare the data with calculations and correct for detector effects, we must also define an event selection using the \enquote{truth} information in the MC event record. This selection uses particle-level objects to match as closely as possible the selection at detector level and is called a \emph{fiducial} particle-level selection. Particle-level events are required to contain exactly one electron and one muon with opposite-sign electric charges and at least two particle-level jets, one of which must contain a $b$-hadron. The \cosphi\ distribution is then constructed from the particle-level top quarks and charged leptons in the same manner as at detector level.
 
The response of the detector, the event selections, and the top-quark reconstruction distort the shape of the \cosphi\ distribution.
The observed distribution is corrected for these effects with a simple method: a simulation-based calibration curve which connects any value at the detector level to the corresponding particle-level value. We correct the data for detector effects by using a unique calibration curve built for each signal and validation region based on the expected signal model, after subtracting the expected contribution from background processes. Due to limited resolution of the reconstructed mass of the \ttbar system, some events that truly belong to the validation regions can enter the signal region at detector level. These events are treated as detector effects.
 
To build these curves, MC event samples are created with alternative values of $D$ by reweighting the events, following the procedure described in Methods~\ref{sec:reweighting}.
The calibration curve corrects the value $D_{\mathrm{detector}}$ measured at the detector level to a corresponding value $D_{\mathrm{particle}}$ at particle level.
To construct the calibration curve, several hypotheses for different values of $D$, denoted by $D^{\prime}_{\mathrm{particle}}$ with a corresponding $D^{\prime}_{\mathrm{detector}}$ value, are created corresponding to changes in the expected value of entanglement.
 
The pairs of $D^{\prime}_{\mathrm{detector}}$ and $D^{\prime}_{\mathrm{particle}}$ are plotted in Figure~\ref{fig:final_results}(a). A straight line interpolates between the points. With this calibration curve, any value for $D_{\mathrm{detector}}$ can be calibrated to the particle level.
 
\begin{figure}[!thbp]
\centering
\includegraphics[width=0.55\textwidth]{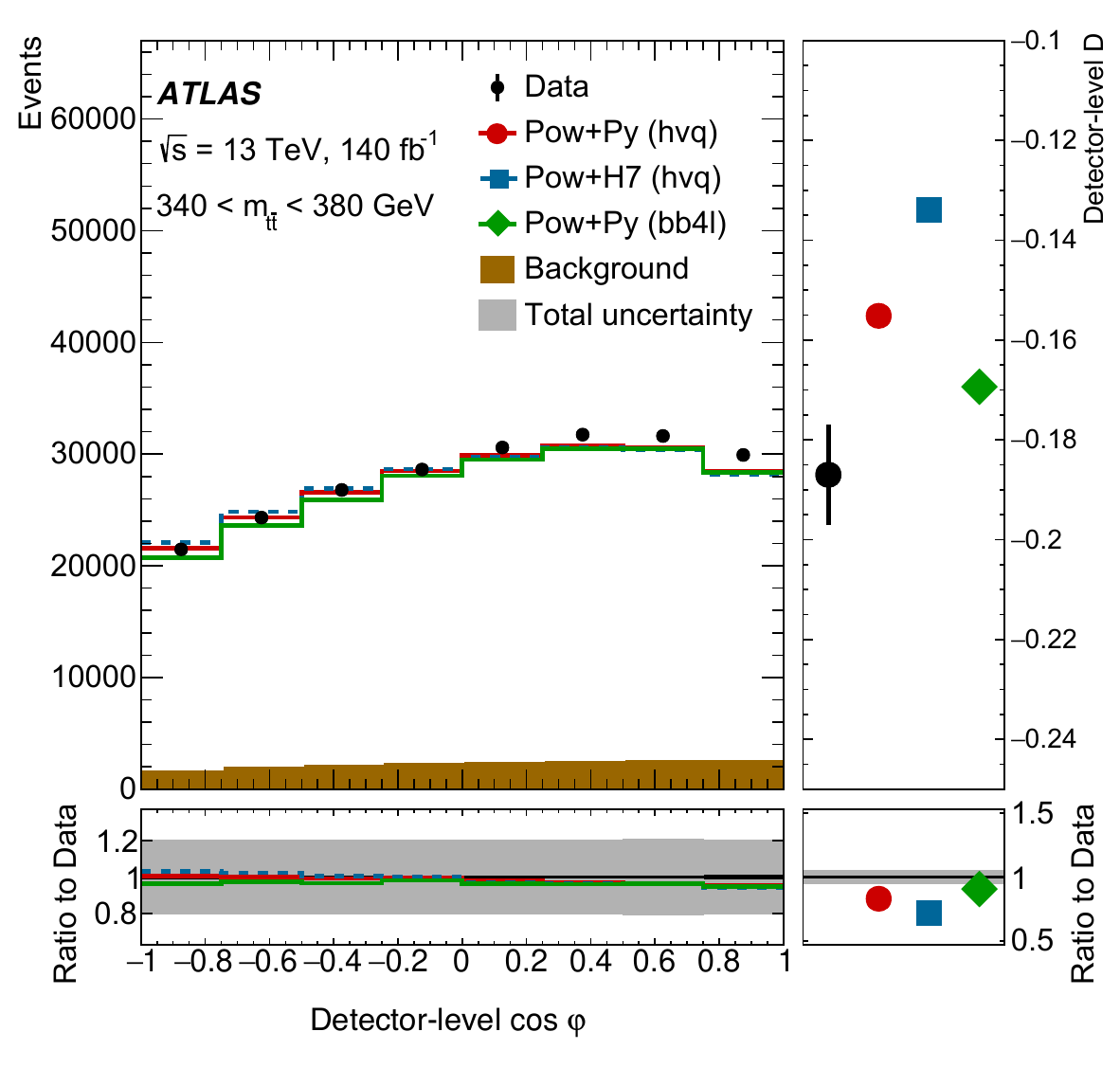}
\caption{The left panel shows the \cosphi observable in the signal region at detector level and the right panel shows the entanglement marker $D$, calculated from the detector-level distributions, from three different MC generators; the \POWPY and \POWHER heavy-quark models, labelled \enquote{Pow+Py (hvq)} and  \enquote{Pow+H7 (hvq)}, respectively, and the \POWPY $bb4\ell$ model, labelled \enquote{Pow+Py ($bb4\ell$)}, are shown after background processes are subtracted. The uncertainty band shows the uncertainties from all sources added in quadrature. The ratios of the predictions to the data are shown at the bottom of the figure. The quoted value for $D$ for the $bb4\ell$ model also includes a subtraction of the single-top-quark background.
}
\label{fig:cosphi}
\end{figure}

Three categories of uncertainties are included in the calibration curves: uncertainties in modeling \ttbar production and decay, uncertainties in modeling the backgrounds, and detector-related uncertainties for both the \ttbar signal and the SM background processes.
Each source of systematic uncertainty can result in a different calibration curve because it changes the shape of the \cosphi distribution at particle level and/or detector level. For each source of systematic uncertainty, the data are corrected using this new calibration curve and the resultant deviation from the data corrected by the nominal curve is taken as the systematic uncertainty of the data due to that source. Systematic uncertainties from all sources are summed in quadrature to determine the final uncertainty in the result.

For all of the detector-related uncertainties, the particle-level quantity is not affected and only detector-level values change.
For signal modeling uncertainties, the effects at particle level propagate to detector level, resulting in shifts in both. Uncertainties in modeling the background processes affect how much background is subtracted from the expected or observed data and can therefore cause changes in the calibration curve. These uncertainties are treated as fully correlated between the signal and background (i.e.\ if a source of systematic uncertainty is expected affect both the signal and background processes, this is estimated simultaneously and not separately).
 
A summary of the different sources of systematic uncertainty and their impact on the result is given in Table~\ref{tab:syst_summary}. The size of each systematic uncertainty depends on the value of $D$ and is given in Table~\ref{tab:syst_summary} for the SM prediction, calculated with \POWPY.
The systematic uncertainties considered in the analysis are described in detail in Method~\ref{sec:systematics}.

To compare the particle-level result with the parton-level entanglement limit $D < -1/3$, the limit must be folded to the particle level.
A second calibration curve is constructed to relate the value of $D_{\mathrm{parton}}$ to the corresponding $D_{\mathrm{particle}}$. The definitions of parton-level top quarks and leptons in the MC generator follow Ref.~\cite{TOPQ-2016-10} and correspond approximately to those of stable top quarks and leptons in a fixed-order calculation.
Only systematic uncertainties related to the modeling of the \ttbar production and decay process are considered when building this calibration curve.
The migration of parton level events from the signal region into the validation regions at particle level and vice versa is very small.

\begin{table}
\centering
\begin{tabular}{l c c  c c}
\toprule
Source of uncertainty & $\Delta D_{\text{observed}} (D=-0.537)$ & $\Delta D$ [\%] & $\Delta D_{\text{expected}} (D=-0.470)$  & $\Delta D$ [\%]  \\
\midrule
Signal modeling  &   $ 0.017$   &   $ 3.2$   &   $0.015$ & $3.2$ \\[0.1cm]
Electrons        &   $ 0.002$   &   $ 0.4$   &   $0.002$ & $0.4$ \\[0.1cm]
Muons            &   $ 0.001$   &   $ 0.2$   &   $0.001$ & $0.1$ \\[0.1cm]
Jets             &   $ 0.004$   &   $ 0.7$   &   $0.004$ & $0.8$ \\[0.1cm]
$b$-tagging      &   $ 0.002$   &   $ 0.4$   &   $0.002$ & $0.4$ \\[0.1cm]
Pile-up          & \phantom{0}$< 0.001$\phantom{$<0$}   & \phantom{0}$< 0.1$\phantom{$<0$}   & \phantom{0}$< 0.001$\phantom{$<0$} & \phantom{0}$< 0.1$\phantom{$<0$} \\[0.1cm]
\MET             &   $ 0.002$   &   $ 0.4$   &   $0.002$ & $0.4$ \\[0.1cm]
Backgrounds      &   $ 0.005$   &   $ 0.9$   &   $0.005$ & $1.1$ \\[0.1cm]
\midrule
Total statistical uncertainty & $ 0.002$ & $ 0.3$ & $ 0.002$ & $ 0.4$\\[0.1cm]
Total systematic uncertainty  & $ 0.019$ & $ 3.5$ & $ 0.017$ & $ 3.6$\\[0.1cm] 
\midrule
Total uncertainty & $ 0.019$ & $ 3.5$ & $ 0.017$ & $ 3.6$\\[0.1cm]
\bottomrule
\\
\end{tabular}
\caption{A summary of the effect of the groups of uncertainties at the expected SM value of $D_{\text{expected}}=-0.470$, corresponding to the \POWPY modeling, and the observed value $D_{\text{observed}}=-0.537$, both in the signal region. \MET denotes the magnitude of the missing transverse momentum. The total systematic uncertainty is calculated as the sum in quadrature of the individual groups of systematic uncertainties.}
\label{tab:syst_summary}
\end{table}

\subsection{Results}
\label{sec:result}

The calibration procedure is performed in the signal region and the two validation regions to correct the data to a \emph{fiducial} phase space at particle level, as described in Section~\ref{sec:analysis}.
All systematic uncertainties are included in the three regions.
The observed (expected) results are:
\begin{equation*}
D = -0.537~\pm 0.002~\text{[stat.]} \pm 0.019~\text{[syst.]}~(-0.470 \pm 0.002~\text{[stat.]} \pm 0.017~\text{[syst.])~,}
\end{equation*}
in the signal region of  $340 < \mttbar < 380$~\GeV\, and:
\begin{equation*}
D = -0.265~\pm 0.001~\text{[stat.]} \pm 0.019~\text{[syst.]}~(-0.258 \pm 0.001~\text{[stat.]} \pm 0.019~\text{[syst.])~,}
\end{equation*}
\begin{equation*}
D = -0.093~\pm 0.001~\text{[stat.]} \pm 0.021~\text{[syst.]}~(-0.103 \pm 0.001~\text{[stat.]} \pm 0.021~\text{[syst.])~,}
\end{equation*}
in the validation regions of $380 < \mttbar < 500$~\GeV\ and $\mttbar > 500$~\GeV, respectively. The expected values are those predicted by \POWPY.
The calibration curve for the signal region and a summary of the results in all regions are presented in Figure~\ref{fig:final_results}.
 
The observed values of the entanglement marker $D$ are compared with the entanglement limit in Figure~\ref{fig:final_results}(b).
The parton-level bound $D = -1/3$ is converted to a particle-level bound by folding the limit to particle level to better highlight the differences between predictions using different parton shower orderings.
For \POWPY, this yields $-0.322 \pm 0.009$, where the uncertainty includes all uncertainties in the \POWPY model except the parton shower uncertainty (for more details of these uncertainties, see Methods~\ref{sec:systematics}).
Similarly, for \POWHER, with an angular-ordered parton shower, a value of $-0.27$ is obtained.
No uncertainties are assigned in this case since it is merely used as an alternative model.

\begin{figure}
\centering
\subfigure[]{
 
\includegraphics[width=0.45\textwidth,trim={0 1.6cm 0 0}]{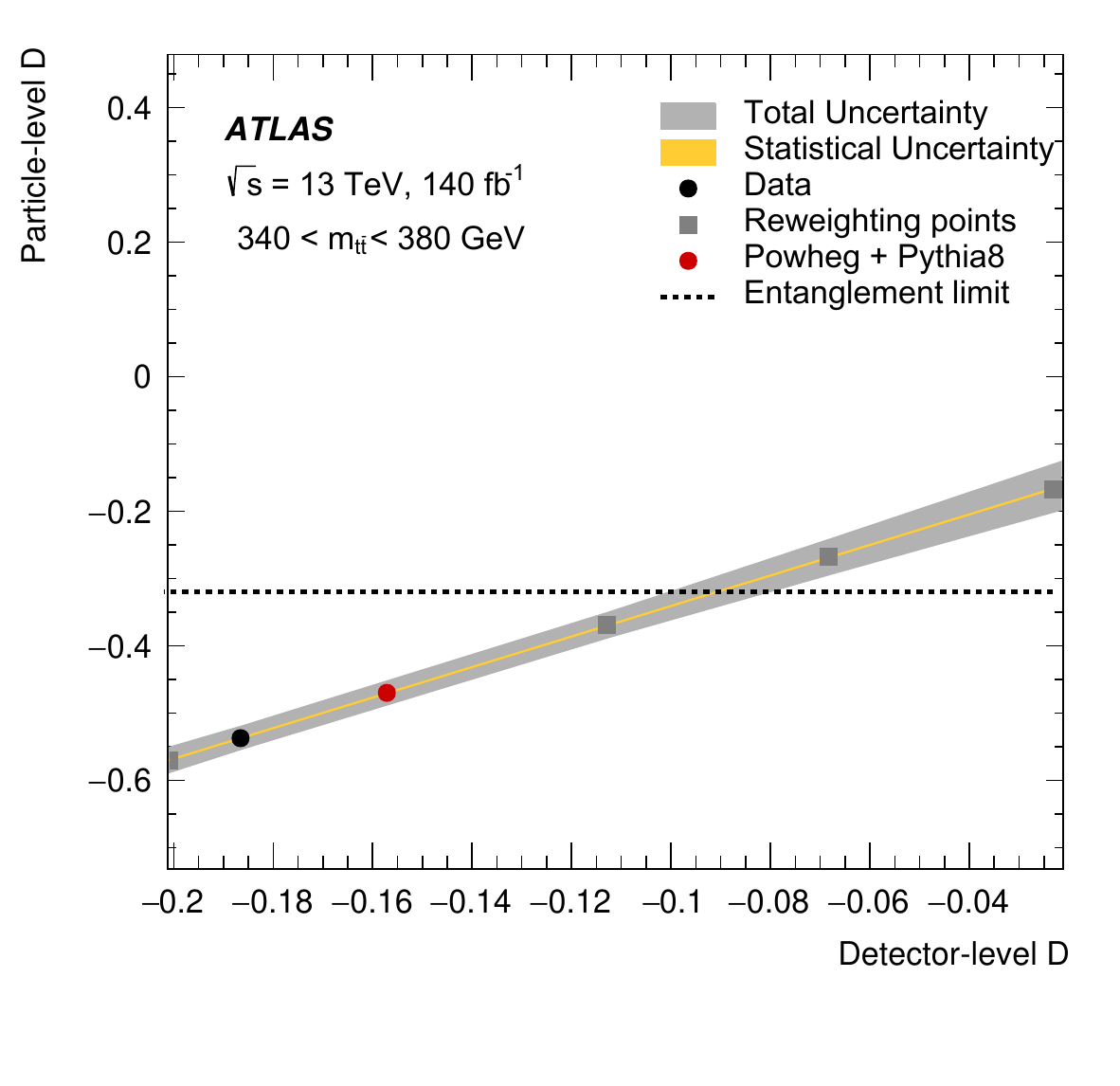}
}
\subfigure[]{
\includegraphics[width=0.52\textwidth]{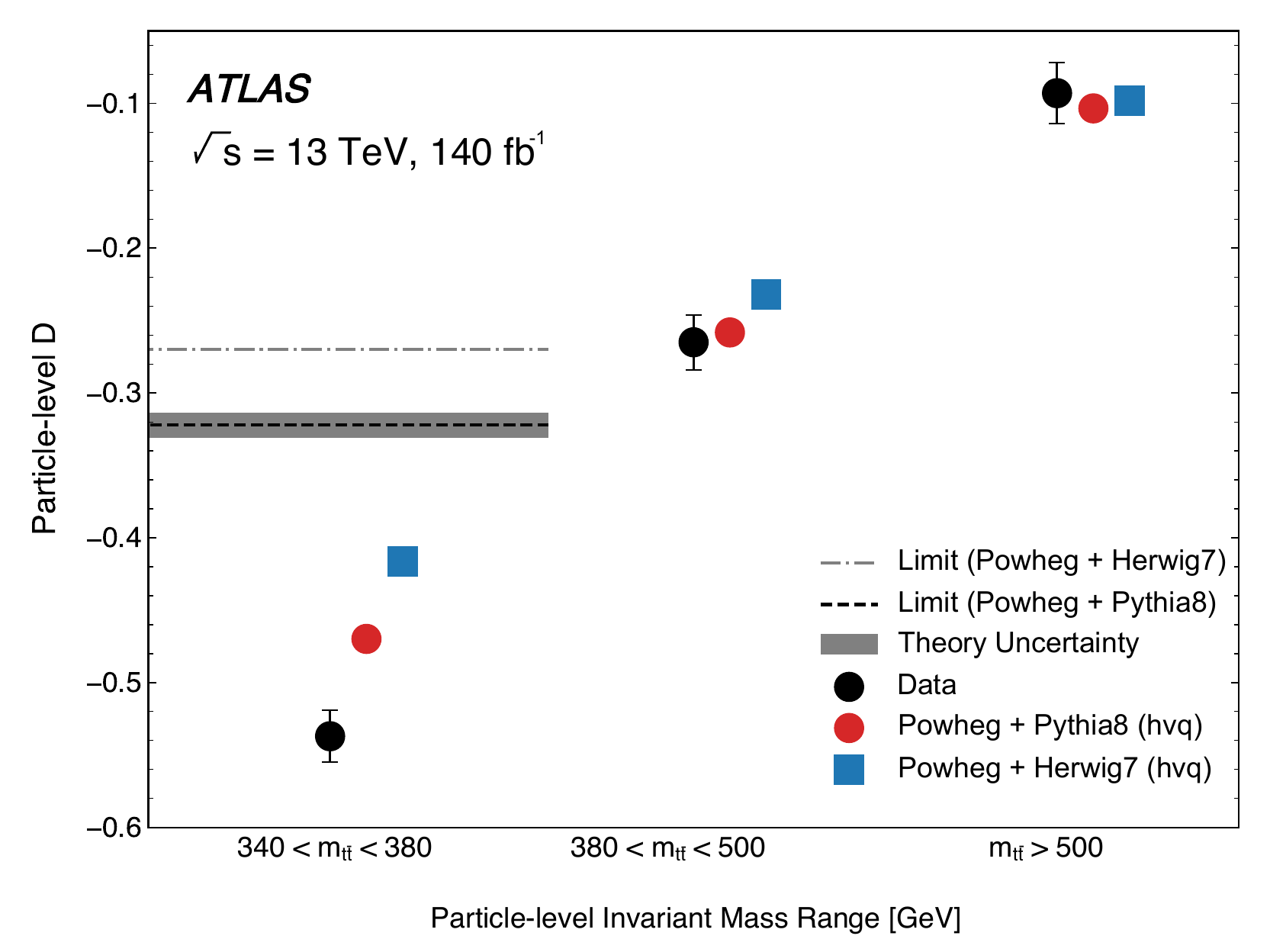}
}

\caption{(a): Calibration curve for the dependence between the particle-level value of $D$ and the detector-level value of $D$, in the signal region.
The yellow band represents the statistical uncertainty, while the grey band represents the total uncertainty obtained by adding the statistical and systematic uncertainties in quadrature.
The measured values and expected values from Powheg + Pythia8 (hvq) are marked with black and red circles, respectively, and the entanglement limit is shown as a dashed line.
(b): The particle-level $D$ results in the signal and validation regions compared with various MC models. The entanglement limit shown is a conversion from its parton-level value of $D = -1/3$ to the corresponding value at particle level, and the uncertainties which are considered for the band are described in the text.
}
\label{fig:final_results}
\end{figure}

\subsection{Discussion}
\label{sec:conc}
 
In both of the validation regions, with no entanglement signal, the measurements are found to agree with the predictions from different MC setups within the uncertainties.
This serves as a consistency check to validate the method used for the measurement.
 
Even though the different models yield different predictions, the current precision of the measurements in the validation regions does not allow us to rule out any of the MC setups that were used.
It is important to note that close to the threshold, non-relativistic QCD processes, such as Coulomb bound state effects, affect the production of \ttbar\ events~\cite{Kiyo:2008bv} and are not accounted for in the MC generators. The main impact of these effects is to change the line-shape of the \mttbar\ spectrum. The impact of these missing effects was tested by introducing them with an ad-hoc reweighting of the MC based on theoretical predictions and the effect was found to be 0.5\%. Other systematic uncertainties on the top-quark decay (1.6\%) and top-quark mass (0.7\%) also change the line-shape in a similar way within our experimental resolution and have a much larger impact. Therefore, the ad-hoc reweighting is not included by default in the measurement since including it would not change the sensitivity of the result within the precision quoted.
 
In the signal region the \POWPY and \POWHER generators yield different predictions.
The size of the observed difference is consistent with changing the method of shower ordering and is discussed in detail in Methods~\ref{appendix:partonshower}.
 
In the signal region, the observed and expected significances with respect to the entanglement limit are well beyond five standard deviations, independently of the MC model used to correct the entanglement limit to account for the \emph{fiducial} phase space of the measurement.
This is illustrated in Figure~\ref{fig:final_results}(b), where the hypothesis of no entanglement is shown. The observed result in the region with $340 < \mttbar < 380$~\GeV\ establishes the formation of entangled \ttbar states.
This constitutes the first observation of entanglement in a quark--antiquark pair.
 
Apart from the fundamental interest in testing quantum entanglement in a new environment, this measurement in top quarks paves the way to use high-energy colliders, such as the LHC, as a laboratory to study quantum information and foundational problems in quantum mechanics.
From a quantum information perspective, high energy colliders are particularly interesting due to their relativistic nature, and the richness of the interactions and symmetries that can be probed there.
Furthermore, highly demanding measurements, such as measuring quantum discord and reconstructing the steering ellipsoid, can be naturally implemented at the LHC due to the vast number of available \ttbar\ events~\cite{Afik:2022dgh}.
From a high-energy physics perspective, borrowing concepts from quantum information theory inspires new approaches and observables that can be used to search for physics beyond the SM~\cite{Aoude:2022imd,Fabbrichesi:2022ovb,Severi:2022qjy,Aoude:2023hxv}.


\appendix
 
\FloatBarrier
\section*{Acknowledgements}


We thank CERN for the very successful operation of the LHC and its injectors, as well as the support staff at
CERN and at our institutions worldwide without whom ATLAS could not be operated efficiently.
 
The crucial computing support from all WLCG partners is acknowledged gratefully, in particular from CERN, the ATLAS Tier-1 facilities at TRIUMF/SFU (Canada), NDGF (Denmark, Norway, Sweden), CC-IN2P3 (France), KIT/GridKA (Germany), INFN-CNAF (Italy), NL-T1 (Netherlands), PIC (Spain), RAL (UK) and BNL (USA), the Tier-2 facilities worldwide and large non-WLCG resource providers. Major contributors of computing resources are listed in Ref.~\cite{ATL-SOFT-PUB-2023-001}.
 
We gratefully acknowledge the support of ANPCyT, Argentina; YerPhI, Armenia; ARC, Australia; BMWFW and FWF, Austria; ANAS, Azerbaijan; CNPq and FAPESP, Brazil; NSERC, NRC and CFI, Canada; CERN; ANID, Chile; CAS, MOST and NSFC, China; Minciencias, Colombia; MEYS CR, Czech Republic; DNRF and DNSRC, Denmark; IN2P3-CNRS and CEA-DRF/IRFU, France; SRNSFG, Georgia; BMBF, HGF and MPG, Germany; GSRI, Greece; RGC and Hong Kong SAR, China; ISF and Benoziyo Center, Israel; INFN, Italy; MEXT and JSPS, Japan; CNRST, Morocco; NWO, Netherlands; RCN, Norway; MEiN, Poland; FCT, Portugal; MNE/IFA, Romania; MESTD, Serbia; MSSR, Slovakia; ARIS and MVZI, Slovenia; DSI/NRF, South Africa; MICINN, Spain; SRC and Wallenberg Foundation, Sweden; SERI, SNSF and Cantons of Bern and Geneva, Switzerland; NSTC, Taipei; TENMAK, T\"urkiye; STFC/UKRI, United Kingdom; DOE and NSF, United States of America.
 
Individual groups and members have received support from BCKDF, CANARIE, CRC and DRAC, Canada; CERN-CZ, PRIMUS 21/SCI/017 and UNCE SCI/013, Czech Republic; COST, ERC, ERDF, Horizon 2020, ICSC-NextGenerationEU and Marie Sk{\l}odowska-Curie Actions, European Union; Investissements d'Avenir Labex, Investissements d'Avenir Idex and ANR, France; DFG and AvH Foundation, Germany; Herakleitos, Thales and Aristeia programmes co-financed by EU-ESF and the Greek NSRF, Greece; BSF-NSF and MINERVA, Israel; Norwegian Financial Mechanism 2014-2021, Norway; NCN and NAWA, Poland; La Caixa Banking Foundation, CERCA Programme Generalitat de Catalunya and PROMETEO and GenT Programmes Generalitat Valenciana, Spain; G\"{o}ran Gustafssons Stiftelse, Sweden; The Royal Society and Leverhulme Trust, United Kingdom.
 
In addition, individual members wish to acknowledge support from CERN: European Organization for Nuclear Research (CERN PJAS); Chile: Agencia Nacional de Investigaci\'on y Desarrollo (FONDECYT 1190886, FONDECYT 1210400, FONDECYT 1230812, FONDECYT 1230987); China: National Natural Science Foundation of China (NSFC - 12175119, NSFC 12275265, NSFC-12075060); Czech Republic: PRIMUS Research Programme (PRIMUS/21/SCI/017); EU: H2020 European Research Council (ERC - 101002463); European Union: European Research Council (ERC - 948254), Horizon 2020 Framework Programme (MUCCA - CHIST-ERA-19-XAI-00), European Union, Future Artificial Intelligence Research (FAIR-NextGenerationEU PE00000013), Italian Center for High Performance Computing, Big Data and Quantum Computing (ICSC, NextGenerationEU), Marie Sklodowska-Curie Actions (EU H2020 MSC IF GRANT NO 101033496); France: Agence Nationale de la Recherche (ANR-20-CE31-0013, ANR-21-CE31-0013, ANR-21-CE31-0022, ANR-22-EDIR-0002), Investissements d'Avenir Idex (ANR-11-LABX-0012), Investissements d'Avenir Labex (ANR-11-LABX-0012); Germany: Baden-Württemberg Stiftung (BW Stiftung-Postdoc Eliteprogramme), Deutsche Forschungsgemeinschaft (DFG - 469666862, DFG - CR 312/5-1); Italy: Istituto Nazionale di Fisica Nucleare (FELLINI G.A. n. 754496, ICSC, NextGenerationEU); Japan: Japan Society for the Promotion of Science (JSPS KAKENHI JP21H05085, JSPS KAKENHI JP22H01227, JSPS KAKENHI JP22H04944, JSPS KAKENHI JP22KK0227); Netherlands: Netherlands Organisation for Scientific Research (NWO Veni 2020 - VI.Veni.202.179); Norway: Research Council of Norway (RCN-314472); Poland: Polish National Agency for Academic Exchange (PPN/PPO/2020/1/00002/U/00001), Polish National Science Centre (NCN 2021/42/E/ST2/00350, NCN OPUS nr 2022/47/B/ST2/03059, NCN UMO-2019/34/E/ST2/00393, UMO-2020/37/B/ST2/01043, UMO-2021/40/C/ST2/00187); Slovenia: Slovenian Research Agency (ARIS grant J1-3010); Spain: BBVA Foundation (LEO22-1-603), Generalitat Valenciana (Artemisa, FEDER, IDIFEDER/2018/048), La Caixa Banking Foundation (LCF/BQ/PI20/11760025), Ministry of Science and Innovation (MCIN \& NextGenEU PCI2022-135018-2, MICIN \& FEDER PID2021-125273NB, RYC2019-028510-I, RYC2020-030254-I, RYC2021-031273-I, RYC2022-038164-I), PROMETEO and GenT Programmes Generalitat Valenciana (CIDEGENT/2019/023, CIDEGENT/2019/027); Sweden: Swedish Research Council (VR 2018-00482, VR 2022-03845, VR 2022-04683, VR grant 2021-03651), Knut and Alice Wallenberg Foundation (KAW 2017.0100, KAW 2018.0157, KAW 2018.0458, KAW 2019.0447); Switzerland: Swiss National Science Foundation (SNSF - PCEFP2\_194658); United Kingdom: Leverhulme Trust (Leverhulme Trust RPG-2020-004); United States of America: U.S. Department of Energy (ECA DE-AC02-76SF00515), Neubauer Family Foundation.


\printbibliography

\clearpage

\begin{refsection}
\section{Methods}
\label{sec:methods}

\subsection{Object Identification in the ATLAS detector}
\label{sec:atlas}
 
ATLAS uses a right-handed coordinate system with its origin at the nominal interaction point in the center of the detector and the $z$-axis along the beam pipe. The $x$-axis points from the interaction point to the center of the LHC ring, and the $y$-axis points upwards. Cylindrical coordinates $(r,\phi)$ are used in the transverse plane, where $\phi$ is the azimuth angle around the $z$-axis. The pseudorapidity is defined in terms of the polar angle $\theta$ as $\eta = - \ln \tan(\theta/2)$. Angular distance is measured in units of $\Delta R \equiv \sqrt{(\Delta\eta)^{2} + (\Delta\phi)^{2}}$.
 
Reconstructed (detector level) objects are defined as follows.
Electron candidates are required to satisfy the \enquote{tight} likelihood-based identification requirement as well as calorimeter- and track- based isolation criteria~\cite{EGAM-2018-01}, and have pseudorapidity $|\eta| < 1.37$ or $1.52 < |\eta| < 2.47$.
Muon candidates are required to satisfy the \enquote{medium} identification requirement as well as track-based isolation criteria~\cite{MUON-2018-03,PERF-2015-10,MUON-2022-01}, and have $|\eta| < 2.5$.
Electrons and muons must have a minimum transverse momentum (\pt) of 25--28~\GeV, depending on the data-taking period.
Showers of particles (jets) that arise from the hadronization of quarks and gluons~\cite{JETM-2018-05} are reconstructed from particle-flow objects~\cite{PERF-2015-09}, using the \antikt algorithm~\cite{Cacciari:2008gp,Fastjet} with a radius parameter $R = 0.4$, a \pt threshold of $25~$\GeV, and a $|\eta|<2.5$ requirement.
Objects can fulfill the criteria for both jet and lepton selection, necessitating the implementation of an overlap removal procedure.
This way, objects are associated with a singular hypothesis.
First, any electron candidates that share a track with a muon candidate are removed. Subsequently, jets within $\Delta R = 0.2$ of an electron are removed, and afterwards, electrons within a region $0.2 < \Delta R < 0.4$ around any remaining jet are rejected. Jets that have fewer than three tracks and are within $\Delta R = 0.2$ of a muon candidate are removed, and muons within $\Delta R = 0.4$ of any remaining jet are discarded.
A Jet-Vertex-Tagger requirement is applied to jets with \pt$ < 60$~\GeV\ and $|\eta| < 2.4$ to suppress jets originating from additional interactions in the same or neighbouring bunch crossings (pile-up)~\cite{PERF-2014-03}.
Jets are tagged as containing $b$-hadrons using the DL1r tagger~\cite{FTAG-2019-07} with a $b$-tagging efficiency of 85\%.
Missing transverse momentum (\ptmiss)~\cite{PERF-2016-07,ATLAS-CONF-2018-023} is determined from the imbalance in the transverse momenta of all reconstructed objects.
 
In order to measure $D$, the top quarks must be reconstructed from their measured decay products.
In the \ttbar dileptonic decay, in addition to charged leptons and jets, there are two neutrinos which are not measured by the detector.
Several methods are available to reconstruct the top quarks from the detector-level charged leptons, jets and \ptmiss. The main method used in this work is the Ellipse method~\cite{Betchart:2013nba}, which is a geometric approach to analytically calculate the neutrino momenta.
This method yields at least one real solution in 85\% of events. We always choose the solution with the lowest top-quark pair invariant mass, to populate the region which is close to threshold. If this method fails (e.g.\ the resultant solutions are all complex), the Neutrino Weighting method~\cite{D0:1997pjc} is used; it assigns a weight to each possible solution by assessing the compatibility of the neutrino momenta and the \ptmiss in the event, after scanning possible values of the pseudorapidities of the neutrinos.
In this analysis, the Neutrino Weighting method is only used in a small fraction of events (about $5\%$).
Furthermore, in Ref.~\cite{TOPQ-2016-10} it was used in all events and the performance was found to be the same between samples that include and exclude spin correlation.
If both methods fail, a simple pairing of each lepton with its closest $b$-tagged jet are used as proxies for the top- and antitop-quark and no attempt is made to reconstruct the neutrinos.
If a second $b$-tagged jet is not present in the event, the leading (highest) \pt untagged jet is used instead.
In all cases, a $W$ boson mass of 80.4~\GeV\ and a top-quark mass of 172.5~\GeV\ are used as input parameters.

In simulated events, parton-level objects are taken directly from the MC history information and are required to have status code of 1, indicating that they are the fundamental particles (partons) of the interaction. Top quarks are required to be partons that decay to a $W$ boson and a $b$ quark, whereas charged leptons are required to be the immediate decay parton from the $W$ boson from the top quark. Particle-level objects are reconstructed using simulated stable particles in the MC simulation before their reconstruction in the detector, but after hadronization.
A particle is defined as stable if it has a mean lifetime greater than 30~ps, within the pseudorapidity acceptance of the detector.
The selection criteria for the particle-level objects are chosen to correspond as closely as possible to the criteria applied to the detector-level objects.
Electrons, muons and neutrinos are required to come from the electroweak decay of a top quark, and are discarded if they arise from the decay of a hadron or a $\tau$-lepton.
Electrons and muons are then \enquote{dressed} by summing their four-momenta with any prompt photons within $\Delta R = 0.1$. Electrons and muons must also be well separated from jet activity. If they lie within $\Delta R < 0.4$ from a jet they are removed from the event. Leptons are also required to have $\pt > 10$~\GeV\ and $|\eta| < 2.5$, and at least one lepton must have $\pt > 25$~\GeV.
Jets are built by clustering all stable particles, using the \antikt algorithm with a radius parameter of $R = 0.4$, and are tagged as containing $b$-hadrons if they have at least one ghost-matched $b$-hadron~\cite{Cacciari:2008gn,Cacciari:2007fd} with $\pt > 5$~\GeV.
Jets are also required to have $\pt > 25$~\GeV\ and $|\eta| < 2.5$.
Each $W$ boson is reconstructed by combining an available electron and electron neutrino or muon and muon neutrino.
The top quark and antitop quark are reconstructed by pairing the two leading $b$-tagged jets, or the $b$-tagged jet and the highest-\pT untagged jet in events with only one $b$-tag, with the reconstructed $W$ bosons.
Both potential jet--lepton combinations are formed and the one which minimizes $|m_t - m(W_1 + b_{1/2})| + |m_t - m(W_2 + b_{2/1})|$ is taken as the correct pairing, where $m_t$ denotes the mass of the top quark, $b_{1/2}$ denotes the two jets selected for the reconstruction, $W_{1/2}$ refers to the reconstructed $W$ bosons, and $m$ is the invariant mass of the objects in brackets.

\subsection{Monte Carlo Simulation}
\label{sec:MonteCarlo}
 
The production of \ttbar events was modeled using the \POWHEGBOX[v2]~heavy-quark (hvq)~\cite{Frixione:2007nw,Nason:2004rx,Frixione:2007vw,Alioli:2010xd} event generator. This generator uses matrix elements calculated at next-to-leading-order (NLO) precision in a strong coupling constant power expansion in QCD with the \NNPDF[3.0nlo]~\cite{Ball:2014uwa} parton distribution function (PDF) set and the \hdamp parameter\footnote{The
\hdamp parameter is a resummation damping factor and one of the
parameters that controls the matching of \POWHEG matrix elements to
the parton shower and thus effectively regulates the
high-\pt radiation against which the \ttbar system recoils.}
set to 1.5\,$m_t$~\cite{ATL-PHYS-PUB-2016-020}.
The decays of the top quarks, including their spin correlations, were modeled at leading-order (LO) precision in QCD.
As an alternative, the \POWHEGBOXRES~\cite{Jezo:2015aia,Jezo:2016ujg} event generator, developed to treat decaying resonances within the \POWHEGBOX framework and including off-shell and non-resonant effects in the matrix element calculation, was used to produce an additional event sample, labelled as $bb4\ell$ in the following.\footnote{Although $bb4\ell$ is the higher-precision MC sample, it cannot be compared directly with the data after they are corrected for detector effects as it is not possible to remove its off-shell component in a formally correct way. However, the effect of using this model was tested in an approximate way and was found to not significantly change the conclusions of the measurement.}
 
In the $bb4\ell$ event sample, spin correlations are calculated at NLO, and full NLO accuracy in \ttbar production and decays is attained.
To model the parton shower, hadronization, and underlying event, the events from both \POWHEGBOX[v2] and \POWHEGBOXRES were interfaced to \PYTHIA[8.230]~\cite{Sjostrand:2014zea}, with parameters set according to the A14 set of tuned parameters~\cite{ATL-PHYS-PUB-2014-021} and using the \NNPDF[2.3lo] set of PDFs~\cite{Ball:2012cx}.
Similarly, the events from \POWHEGBOX[v2]~(hvq) were also interfaced to \HERWIG[7.2.1]~\cite{Bahr:2008pv,Bellm:2019zci}, using the \HERWIG[7.2.1] default set of tuned parameters.
The decays of bottom and charm hadrons were performed by \EVTGEN[1.6.0]~\cite{Lange:2001uf}.
The spin information from the matrix element calculation is not passed to the parton shower programs and therefore is not fully preserved during the shower.

All simulated event samples include pile-up interactions, and the events are reweighted to reproduce the observed distribution of the average number of collisions per bunch crossing.

\subsection{Reweighting the \cosphi distribution}
\label{sec:reweighting}
 
In order to construct the calibration curve, templates for alternative scenarios with different degrees of entanglement, and therefore with different values of $D$, must be extracted.
The degree of entanglement is intrinsic in the calculations of the MC event generators.
However, the effects of entanglement can be directly accessed via $D$, measured from the average of the \cosphi distribution in the event.
Therefore, an event-by-event reweighting based on $D$ is used to vary the degree of entanglement.
Although the measurement uses detector-level and particle-level objects, the observable $D$ is changed at parton level, where it is directly related to the entanglement between the top and antitop spins.
Therefore, each event is reweighted according to its parton-level values of $m_{t\bar t}$ and \cosphi, as described below.

The entanglement marker $D$ is extracted at parton level from the \cosphi distribution by using either the mean of the distribution $D =~-3\cdot \langle \cos\varphi \rangle$ or the slope of the normalized differential cross-section 
$(1/\sigma)\,\dif\sigma/\dif\cos\varphi = (1/2) (1-D\cos\varphi)$.
 
For simplicity, the analysis always uses the mean of the distribution, although the two methods are equivalent.
Thus for the purpose of reweighting, one must change the slope of the \cosphi distribution at parton level.
Each event is reweighted according to this slope, which in turn changes the distributions at particle level and detector level.
 
The observable $D$ depends on the invariant mass of the \ttbar system, $m_{t\bar t}$.
To perform the reweighting, the differential value of $D$ per mass unit as a function of $m_{t\bar{t}}$, $D_\Omega(m_{t \bar t})$, has to be calculated. This is achieved by fitting a third-order polynomial of the form:
\begin{equation*}
D_\Omega(m_{t\bar t}) = x_0 + x_1 \cdot m_{t\bar t}^{-1} + x_2 \cdot m_{t\bar t}^{-2} + x_3 \cdot m_{t\bar t}^{-3}~,
\end{equation*}
where $x_0,x_1,x_2,x_3$ are constants.
This parametrization was found to describe well the value of $D_\Omega(m_{t\bar t})$, in good agreement with the MC prediction.
The values of the parameters of $D_\Omega(m_{t\bar t})$ depend on the MC event generator and have to be calculated for the nominal sample and for the effect of each of the $t\bar{t}$ theory systematic uncertainties, since they change the parton-level \cosphi values and thus $D_\Omega(m_{t\bar t})$.

The reweighting method is a simple scaling of the \cosphi distribution according to the desired new value of $D$. This is done by assigning a weight $w$ to each event at parton level as:
\begin{equation*}
w = \frac{1 - D_\Omega(m_{t \bar t}) \cdot \mathcal{X} \cdot \cos \varphi}{1 - D_\Omega(m_{t \bar t}) \cdot \cos \varphi}~,
\end{equation*}
with $\mathcal{X}$ as the scaling hypothesis of $D$. If, for example, $\mathcal{X}=1.2$, it means that $D$ is scaled up by 20\% relative to its nominal value.
In order to build the calibration curve, four alternative values of $D$ are considered, with $\mathcal{X} = 0.4, 0.6, 0.8, 1.2$, in addition to the nominal value without reweighting ($\mathcal{X} = 1.0$).
It is important to note that these $\mathcal{X}$ values change $D$ across the entire $m_{t\bar{t}}$ spectrum.
In Figure~\ref{fig:cosphi_reweighting} the parton-level distribution of $D$ is shown in the signal region before and after reweighting.
 
\begin{figure}[!thbp]
\centering
\includegraphics[width=0.6\textwidth]{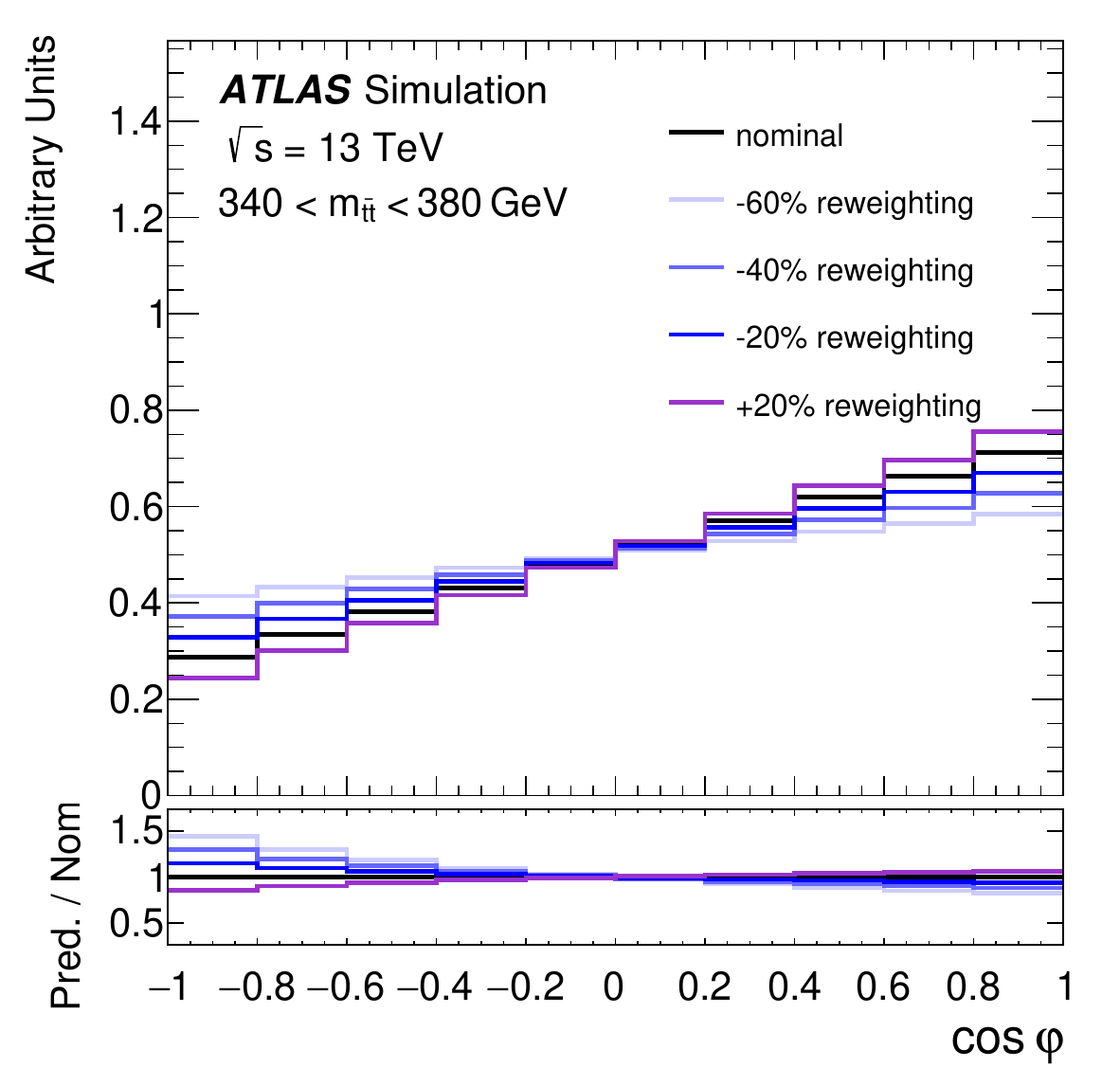}
\caption{Example of the nominal \cosphi distribution and the results of applying the reweighting technique with $\mathcal{X} = 0.4, 0.6, 0.8, 1.2$ in the signal region at parton level. The lower panel shows the ratio of each $D$ value after reweighting (\enquote{Pred.}) to the nominal $D$ value (\enquote{Nom.}).}
\label{fig:cosphi_reweighting}
\end{figure}

\subsection{Background modeling}

Simulated data in the form of MC samples were produced using either the full ATLAS detector simulation~\cite{SOFT-2010-01} based on the \GEANT framework~\cite{Agostinelli:2002hh} or, for the estimation of some of the systematic uncertainties, a faster simulation with parameterized showers in the calorimeters~\cite{ATL-PHYS-PUB-2010-013}. The effect of pile-up was modeled by overlaying each hard-scattering event with inelastic $pp$ collisions generated with \PYTHIA[8.186]~\cite{Sjostrand:2007gs} using the \NNPDF[2.3lo] set of PDFs~\cite{Ball:2012cx} and the A3 set of tuned parameters~\cite{ATL-PHYS-PUB-2016-017}. Except for the events simulated with \SHERPA, the \EVTGEN program was used to simulate bottom and charm hadron decays. If not mentioned otherwise, the top-quark mass was set to $m_{t} = 172.5\,\GeV$. All event samples that were interfaced with \PYTHIA used the A14 set of tuned parameters~\cite{ATL-PHYS-PUB-2014-021} and the \NNPDF[2.3lo] PDF set.
 
Single-top quark $tW$ associated production was modeled using the
\POWHEGBOX[v2]~\cite{Re:2010bp,Nason:2004rx,Frixione:2007vw,Alioli:2010xd}
event generator, which provides matrix elements at NLO in the strong coupling constant \alphas\ in the five-flavor
scheme with the \NNPDF[3.0nlo]~\cite{Ball:2014uwa} PDF set.  The functional form of the
renormalization and factorization scales was set to the default scale,
which is equal to the top-quark mass.
The diagram-removal scheme~\cite{Frixione:2008yi} was employed to handle the interference
with \ttbar production~\cite{ATL-PHYS-PUB-2016-020}.
The inclusive cross-section was corrected to the theory prediction
calculated at NLO in QCD with next-to-next-leading-logarithm (NNLL) soft-gluon
corrections~\cite{Kidonakis:2010ux,Kidonakis:2013zqa}.  For $pp$
collisions at a center-of-mass energy of $\rts=\SI{13}{\TeV}$, this
cross-section corresponds to $\sigma(tW)_\text{NLO+NNLL}= 71.7\pm3.8$\,pb.  The uncertainty in
the cross-section due to the PDF was estimated using the \MSTW[nnlo] 90\%
CL~\cite{Martin:2009iq,Martin:2009bu} PDF set, and was added in
quadrature to the effect of the scale uncertainty.

Samples of diboson final states ($VV$), where $V$ denotes a $W$ or $Z$ boson, were simulated with the
\SHERPA[2.2.2]~\cite{Bothmann:2019yzt} event generator,
including off-shell effects and Higgs boson contributions, where appropriate.
Fully leptonic final states and semileptonic final states, where one boson
decays leptonically and the other hadronically, were generated using
matrix elements at NLO accuracy in QCD for up to one additional parton
and at LO accuracy for up to three additional parton
emissions. Samples for the loop-induced processes $gg \to VV$ were
generated using LO-accurate matrix elements for up to one
additional parton emission for both the cases of fully leptonic and
semileptonic final states. The matrix element calculations were matched
and merged with the \SHERPA parton shower based on Catani--Seymour
dipole factorization~\cite{Gleisberg:2008fv,Schumann:2007mg} using the MEPS@NLO
prescription~\cite{Hoeche:2011fd,Hoeche:2012yf,Catani:2001cc,Hoeche:2009rj}.
The virtual QCD corrections were provided by the
\OPENLOOPS library~\cite{Buccioni:2019sur,Cascioli:2011va,Denner:2016kdg}. The
\NNPDF[3.0nnlo] set of PDFs was used~\cite{Ball:2014uwa}, along with the
dedicated set of tuned parton-shower parameters developed by the
\SHERPA authors.
 
The production of $V+$\,jets events was simulated with the
\SHERPA[2.2.11]~\cite{Bothmann:2019yzt}
event generator using NLO matrix elements for up to two partons, and LO matrix elements
for up to five partons, calculated with the Comix~\cite{Gleisberg:2008fv}
and \OPENLOOPS[2]~\cite{Buccioni:2019sur,Cascioli:2011va,Buccioni:2017yxi,Denner:2016kdg} libraries. They
were matched with the \SHERPA parton shower~\cite{Schumann:2007mg} using the \MEPSatNLO
prescription~\cite{Hoeche:2011fd,Hoeche:2012yf,Catani:2001cc,Hoeche:2009rj}.
The set of tuned parameters developed by the \SHERPA authors was used, along with
the \NNPDF[3.0nnlo] set of PDFs~\cite{Ball:2014uwa}.

The production of \ttV events was modeled using the \MGNLO[2.3.3]
\cite{Alwall:2014hca} event generator, which provides matrix elements at NLO in the strong coupling constant \alphas
with the \NNPDF[3.0nlo]~\cite{Ball:2014uwa} PDFs. The functional form of the renormalization and
factorization scales was set to $0.5 \times \sum_i
\sqrt{m^2_i+p^2_{\text{T},i}}$, where the sum runs over all the particles
generated from the matrix element calculation. Top quarks were decayed
at LO using \MADSPIN~\cite{Frixione:2007zp,Artoisenet:2012st} to
preserve spin correlations. The events were interfaced with
\PYTHIA[8.210]~\cite{Sjostrand:2014zea} for the simulation of parton showering and
hadronization.
The cross-sections were calculated at NLO QCD and NLO EW accuracy using
\MGNLO as reported in Ref.~\cite{LHCHiggsCrossSectionWorkingGroup:2016ypw}.
For \ttll events, the cross-section was scaled by
an off-shell correction estimated at one-loop level in \alphas.

The production of \ttH events was modeled using the
\POWHEGBOX[v2]~\cite{Frixione:2007nw,Nason:2004rx,Frixione:2007vw,Alioli:2010xd,Hartanto:2015uka}
event generator, which provides matrix elements at NLO in the strong coupling
constant \alphas in the five-flavor scheme with the \NNPDF[3.0nlo]~\cite{Ball:2014uwa} PDF set.
The functional form of the renormalization and factorization scales was
set to $\sqrt[3]{m_\text{T}(t)\cdot m_\text{T}(\bar{t}) \cdot m_\text{T}(H)}$.
The events were interfaced with \PYTHIA[8.230].
The cross-section was calculated at NLO QCD and NLO EW accuracy using
\MGNLO as reported in Ref.~\cite{LHCHiggsCrossSectionWorkingGroup:2016ypw}. The predicted value at \rts~=~\SI{13}{\TeV} is
507$^{+35}_{-50}$~fb, where the uncertainties were estimated from
variations of both \alphas and the renormalization and factorization scales.

The background from non-prompt or fake leptons was modeled using simulated MC events to describe the shape of the kinematic distributions. MC event generator information is used to distinguish events with prompt leptons from events with non-prompt or fake leptons.  The normalization of this background was obtained from data by using a dedicated control region. This control region uses the same basic event selection as the signal and validation regions, the only difference being that the electric charges of the electron and muon must have the same sign.
Within this control region, the number of simulated prompt-lepton events is subtracted from the observed number of data events. The number of events remaining is then divided by the number of simulated fake-lepton events, resulting in a normalization factor of 1.4.  This scale factor is then applied to the simulated fake-lepton events in the signal and validation regions.

\subsection{Systematic uncertainties}
\label{sec:systematics}
 
The systematic uncertainties can be divided into three separate categories: signal modeling uncertainties, which stem from the theory prediction of \ttbar production; object systematic uncertainties, which arise from the uncertainty in the detector response to objects used in the analysis; and background modeling systematic uncertainties, which are related to the theory prediction of the SM backgrounds.
All systematic uncertainties, grouped according to their sources, are described in the following sections.
The signal modeling uncertainties were found to dominate the overall uncertainty of this measurement.

For each source of systematic uncertainty, a new calibration curve is created and the simulated (or observed) data are corrected, resulting in a shifted corrected result.  In most cases the systematic uncertainty is taken to be the difference between the nominal expected/observed result and the systematically shifted result.
In cases where a systematic shift only effects the background model (e.g.\ background cross-section uncertainties), the systematically shifted background sample is subtracted from the data instead before the calibration is performed.
In cases where the systematic uncertainty is one-sided, the uncertainty is symmetrized.
In cases where uncertainties are asymmetric, the larger of the two variations is symmetrized. The signal modeling uncertainties dominate the measurement and their estimated sizes are presented in Table~\ref{tab:modeling_systematics_summary_with_parton_main}.

\subsubsection{Signal modeling uncertainties}
 
Signal modeling uncertainties are those related to the choice of \POWHEGBOX + \PYTHIA as the nominal MC setup as well as those affecting the theoretical calculation itself.
These systematic uncertainties are considered in two forms: alternative event generators, and weights. For the alternative-generator uncertainties, the difference between the calibrated values of $D$ is taken as the systematic uncertainty.
For the systematic uncertainties involving weights, the difference between the calibrated $D$ values for the nominal sample and the weight-shifted sample is taken as the uncertainty. These uncertainties follow the description in Ref.~\cite{ATL-PHYS-PUB-2011-005} and are enumerated as follows:
 
\begin{itemize}
 
\item \textbf{pThard setting}: The region of phase space that is vetoed in the showering when matched to a parton shower is varied by changing the internal \texttt{pThard} parameter of \POWHEGBOX from 0 to 1, following the prescription described in Ref.~\cite{Hoche:2021mkv}. 
 
\item \textbf{Top-quark decay}: The uncertainty in the modeling of the decay of the top quarks and of the \mttbar\ line-shape is estimated by comparing the nominal decay in \POWHEGBOX with the decays modeled with \MADSPIN~\cite{Frixione:2007zp,Artoisenet:2012st}. The effect of this uncertainty is to shift the \mttbar\ line-shape to lower or higher values which alters the degree of entanglement entering the signal region. Thus, this is one of the most impactful sources systematic uncertainty.
 
\item \textbf{NNLO QCD + NLO EW reweighting}: The uncertainty due to missing higher-order corrections is estimated by reweighting the \pt of the top quarks, the \pt of the \ttbar system, and the $m_{\ttbar}$ spectra at parton level to match the predicted NNLO QCD and NLO EW differential cross-sections~\cite{Czakon:2015owf,Czakon:2017wor}.
 
\item \textbf{Parton shower and hadronization}: This uncertainty is estimated by comparing two different parton-shower and hadronization algorithms, \PYTHIA and \HERWIG, interfaced to the same matrix element event generator (\POWHEGBOX).
 
\item \textbf{Recoil scheme}:
The nominal sample uses a recoil scheme where partons recoil against $b$-quarks. This recoil scheme changes the modeling of second and subsequent gluon emissions from quarks produced by colored resonance decays, such as the $b$-quark in a top-quark decay, and therefore affects how the momentum is rearranged between the $W$ boson and the $b$-quark. An alternative sample is produced in which the recoil is set to be against the top quark itself for the second and subsequent emissions~\cite{TOPQ-2017-17}.
 
\item \textbf{Scale uncertainties:}
The renormalization and factorization scales are raised and lowered by a factor of 2 in the nominal \POWHEG setup, including simultaneous variations in the same direction. The envelope of results from all of these variations is taken as the final uncertainty.
 
\item \textbf{Initial-state radiation}:
The uncertainty due to initial-state radiation is estimated by choosing the Var3c up/down variations of the A14 tune as described in Ref.~\cite{ATL-PHYS-PUB-2017-007}.
 
\item \textbf{Final-state radiation}:
The impact of final-state radiation is evaluated by doubling or halving the renormalization scale for emissions from the parton shower.
 
\item \textbf{Parton distribution function (PDF)}:
The systematic uncertainty due to the choice of PDF is assessed using the PDF4LHC15 eigenvector decomposition~\cite{Butterworth:2015oua}. The full difference between the results from the nominal PDF and the varied PDF is taken and symmetrised for each of the 30 eigenvectors. The quadrature sum of all result variations is quoted in Table~\ref{tab:modeling_systematics_summary_with_parton_main}.
 
\item \textbf{\hdamp setting}:
The \hdamp parameter is a resummation damping factor and one of the parameters that controls the matching of \POWHEGBOX matrix elements to the parton shower and thus effectively regulates the high-\pt radiation against which the \ttbar system recoils.
The systematic uncertainty due to the chosen value of the \hdamp parameter is assessed by comparing the nominal \POWPY result with one where the \hdamp parameter is increased by a factor of two.
 
\item \textbf{Top-quark mass}:
The effect of the top-quark mass uncertainty is examined by comparing the nominal sample with alternative samples that use $m_t = 172$ or 173~\GeV\ in the simulation.
 
\end{itemize}

\begin{table}[h!]
\centering
\begin{tabular}{l c}
\toprule
Systematic uncertainty source &  Relative size (for SM $D$ value)\\
\midrule
Top-quark decay              & 1.6\% \\
Parton distribution function & 1.2\% \\
Recoil scheme                & 1.1\% \\
Final-state radiation        & 1.1\% \\
Scale uncertainties          & 1.1\% \\
NNLO QCD + NLO EW reweighting             & 1.1\% \\
pThard setting              & 0.8\% \\
Top-quark mass                     & 0.7\% \\
Initial-state radiation      & 0.2\% \\
Parton shower and hadronization & 0.2\% \\
\hdamp setting               & 0.1\% \\
\bottomrule
\end{tabular}
\vspace{2mm}
\caption{Relative sizes of the signal modeling uncertainties at the SM expectation point $D_{\textrm{particle}} = -0.47$ for the nominal \POWHEGBOX sample.}
\label{tab:modeling_systematics_summary_with_parton_main}
\end{table}
 
\subsubsection{Object systematic uncertainties}
 
Systematic uncertainties which originate from the uncertainty in the detector response to the objects used in the analysis are estimated.
 
\begin{itemize}
\item \textbf{Electrons:} The systematic uncertainties considered for electrons arise mainly from uncertainties in their trigger, reconstruction, identification, and isolation efficiencies, and are estimated using tag-and-probe measurements in $Z$ and $J/\psi$ decays~\cite{EGAM-2018-01,PERF-2017-01}.
Electron-related systematic uncertainties have a negligible impact on the final measurement, with a total contribution of about 0.2\%.
 
\item \textbf{Muons:} The systematic uncertainties considered for muons arise from uncertainties in their trigger, identification, and isolation efficiencies, and their energy scale and resolution, and are estimated using tag-and-probe measurements in $Z$ and $J/\psi$ decays~\cite{MUON-2018-03,PERF-2015-10,MUON-2022-01}.
Muon-related systematic uncertainties have a negligible impact on the final measurement, with a total contribution of about 0.3\%.
 
\item \textbf{Jets:} The systematic uncertainties associated with jets are separated into those related to the jet-energy scale and resolution (JES and JER)~\cite{JETM-2018-05} and those related to the jet-vertex tagger (JVT) algorithm~\cite{PERF-2014-03}.
The JES (JER) uncertainty consists of 31 (13) individual components that are added in quadrature with the JVT uncertainty to obtain the total jet uncertainty. The largest contribution from a single source is 0.2\%.
 
\item \textbf{$b$-tagging:} The estimation of these uncertainties is described in Ref.~\cite{FTAG-2018-01}. A total of 17 independent systematic variations are considered: 9 related to $b$-hadrons, 4 related to $c$-hadrons, and 4 related to light-jet misidentification. In addition, two high-\pt extrapolation uncertainties are taken into account. The largest contribution from a single systematic variation is 0.4\%.
 
\item \textbf{\MET:} All object-based uncertainties are fully correlated with the reconstruction of the event's \MET object, the magnitude of the  \ptmiss vector. However, there are some uncertainties specific to the reconstruction of \MET which concern soft tracks not matched to leptons or jets.
These uncertainties are divided into parallel and perpendicular response components as well as a scale uncertainty~\cite{PERF-2016-07}.
These have a negligible effect on the measurement.
 
\item \textbf{Pile-up:} The effect of pile-up was modeled by overlaying the simulated hard-scattering events with inelastic $pp$ events.
In order to assess the systematic uncertainty due to pile-up, the reweighing performed to match simulation to data is varied within its uncertainty~\cite{PERF-2014-03}.
The resulting uncertainty has an effect of less than 0.1\%.
 
\item \textbf{Luminosity:} The luminosity uncertainty only changes the normalization of the signal and background samples. The value of $D$ is calculated from the normalized \cosphi distribution and therefore is not affected by varying the sample normalization. However, the total expected statistical uncertainty can be affected by the luminosity uncertainty. This analysis uses the latest integrated luminosity estimate of $140.1 \pm 1.2$~fb$^{-1}$~\cite{DAPR-2021-01}. Its uncertainty affects the measurement by less than 0.1\%.

\end{itemize}

\subsubsection{Background modeling systematic uncertainties}
Background events are a relatively small source of uncertainty in this measurement since the event selection and top-quark reconstruction, especially the $m_{t\bar{t}}$ constraint, tend to suppress them.
The uncertainties and their sources are listed in the following.
 
\begin{itemize}
 
\item \textbf{Single top quark:}
Two uncertainties are considered for the single-top quark background: a cross-section uncertainty of 5.3\% based on the NNLO cross-section uncertainty~\cite{Kidonakis:2013zqa}, and an uncertainty for the choice of scheme used to remove higher-order diagrams that overlap with the $t\bar{t}$ process.
For the latter, the nominal \POWPY sample, generated with the diagram-removal scheme~\cite{Frixione:2008yi}, was compared with an alternative
sample generated using the diagram-subtraction scheme~\cite{Frixione:2008yi,ATL-PHYS-PUB-2016-020}.
The cross-section uncertainty has a 0.4\% effect on the measurement, whereas the choice of diagram scheme has less than a 0.1\% effect on the measurement.
 
\item \textbf{$t\bar{t}+X$:}
A normalization uncertainty is considered for each of the $t\bar{t}+X$ backgrounds: a cross-section uncertainty of $^{+10\%}_{-12\%}$ for $t\bar{t}+Z$, and $^{+13\%}_{-12\%}$ for $t\bar{t}+W$. Both are based on the NLO cross-section uncertainty derived from renormalization and factorization scale variations and PDF uncertainties in the matrix element calculation. These uncertainties have a negligible effect on the measurement, since the $t\bar{t}+X$ processes make a very small contribution in the signal region.
 
\item \textbf{Diboson:}
A normalization uncertainty of $\pm 10\%$ is considered for the diboson process to account for the difference between the NLO precision of the \Sherpa event generator and precision of the theoretical cross-sections calculated to NNLO in QCD with NLO EW corrections. This simple $K$-factor approach is taken, rather than a more elaborate prescription, because the diboson background is small and the phase space selected by the analysis ($\mttbar < 380$~\GeV) is unlikely to be sensitive to shape effects in the EW corrections, typically observed in high-\pt tails. This uncertainty has less than a 0.1\% effect on the measurement.
 
\item \textbf{$Z\rightarrow\tau\tau$:}
A conservative cross-section uncertainty of $\pm 20\%$ is applied to the $Z\rightarrow\tau\tau$ background in order to account for the uncertainty in the cross-section prediction (which is much smaller than this variation) as well as to account for some mismodeling of the rate of associated heavy-flavor production, which is typically seen in $ee$ and $\mu\mu$ dileptonic $t\bar{t}$ analyses and was estimated to be a 5\% (3\%) effect in previous iterations of this analysis that included the $ee$ ($\mu\mu$) channel. This assumption is conservative as it is not possible to isolate a pure $Z\rightarrow\tau\tau$ control region in which to estimate this effect, and therefore additional lepton-flavor-related effects present in the $ee$ and $\mu\mu$ channels are also being included. This uncertainty has a noticeable impact on the final measurement, becoming the largest background-related uncertainty. It becomes large, despite this background being relatively small, because the reconstruction-level $Z\rightarrow\tau\tau$ \cosphi distribution is quite flat and therefore subtracting even a relatively small amount of $Z\rightarrow\tau\tau$ background can noticeably affect the mean of the overall \cosphi distribution and therefore the $D$ observable.
This uncertainty has an impact of 0.8\% on the measurement.

\item \textbf{Fake and non-prompt leptons:}
A normalization uncertainty of $\pm 50\%$ is assigned to account for the uncertainty in the total yield of fake or non-prompt leptons in the signal region compared to the same-sign control region in order to ensure adequate coverage for our understanding of the rates of these types of events. It is a conservative uncertainty based on the observed level of data--MC agreement in the same-sign region. The uncertainty has only a 0.1\% effect on the final measurement.
 
\end{itemize}
 
The majority of systematic uncertainties that are considered are inconsequential to the measurement, and the dominant systematic uncertainties arise mostly from the signal modeling and the $Z \to \tau\tau$ cross-section uncertainty. These findings are true for the validation regions as well.

\subsection{Parton shower and hadronization effects}
\label{appendix:partonshower}
 
The studies described in the following were performed to gain a more detailed understanding of why the different parton-shower and hadronization algorithms yield different values for the entanglement- and spin-correlation-related observables.
The nominal MC sample was produced with the NLO matrix element implemented in \POWHEGBOX (hvq). The four-momenta produced with \POWHEGBOX
were interfaced with either \PYTHIA or \HERWIG for the parton shower, hadronization and underlying-event model.

At parton level, the two predictions are nearly identical, while at the stable-particle and detector levels the two predictions show larger differences in the shape of the \cosphi distributions. A parton-level measurement would therefore suffer from the ambiguity in \cosphi, while the particle-level measurement presented in this paper does not. An extensive suite of studies was performed to understand the origin of this difference.
 
Apart from using different parameter-tuning strategies, there are two main differences between the two parton-shower algorithms: their hadronization model and the shower ordering.
While \PYTHIA is based on the Lund string model and uses a \pt-ordered dipole shower~\cite{Gustafson:1994cd,Lonnblad:1995yk,Friberg:1996xc}, the \HERWIG samples used in this study are based on a cluster model and uses an angular-ordered shower as the default~\cite{Webber:1997iw}.

A comparison between MC simulations with different hadronization models was performed.
For one study, \SHERPA was used with either a string or a cluster model for hadronization.
For the other study \HERWIG[7] was used, again comparing the effects of using either a string or a cluster model.
Changing the hadronization model has shown in both cases to have a negligible
effect on the \cosphi distribution, both when not placing a cut on $m_{t\bar{t}}$ and when using a smaller part of phase space close to the signal region of the analysis, with $\mttbar < 380$~\GeV.
Instead, most of the differences seem to originate from the different orderings in the parton shower.
To illustrate this, different event generator setups were used for simulation and the corresponding \cosphi distributions were compared at particle level.
The \cosphi distributions for the \POWPY and \POWHER samples used in the analysis are shown in Figure~\ref{fig:PSHAD_Direct3a}, together with distributions for two different setups of \HERWIG[7] in Figure~\ref{fig:PSHAD_Direct3b}.
In these setups, \HERWIG[7] was used both for the production of the \ttbar events and for the parton shower, hadronization and underlying event.
The samples were produced at LO, using either a dipole shower or an angular-ordered shower. All distributions are normalized to unity.
A difference of up to 6\% is observed when examining the ratio of \POWHER to \POWPY distributions. The same behavior is observed when comparing the two different showering orders for \HERWIG.
 
The similarities between the samples used in this analysis and the \HERWIG samples with different showering orders implies that the ordering of the shower is the main cause of the observed differences. It has to be noted, however, that \POWHEG\ does not pass the spin correlation information to the parton shower algorithms, while this is done in the LO \HERWIG\ setup used to study these hadronization effects.
 
These findings lead to the conclusion that performing the measurement at particle level is more attractive, since the difference in the predictions when extrapolating from parton to particle level can be isolated and not taken as full systematic uncertainty. In the validation regions, the level of agreement between either \POWPY or \POWHER and the data is similar. Since the measurement is performed at the stable-particle level, the parton-level prediction for the entanglement limit was folded to the particle level as well, using a special calibration curve for this step. The prediction for the entanglement limit with \POWHER is further away from the data measurement than the one for \POWPY. This difference is not symmetrized. All uncertainties in the \POWPY prediction itself are folded to particle level as well and are included in the grey uncertainty band in Figure~\ref{fig:final_results}.

The procedure used in MC event generators to combine the matrix element with a parton-shower algorithm requires special attention in future higher-precision quantum information studies at the LHC.
 
\begin{figure}[!htb]
\centering
\subfigure[]{
\includegraphics[width=0.435\textwidth]{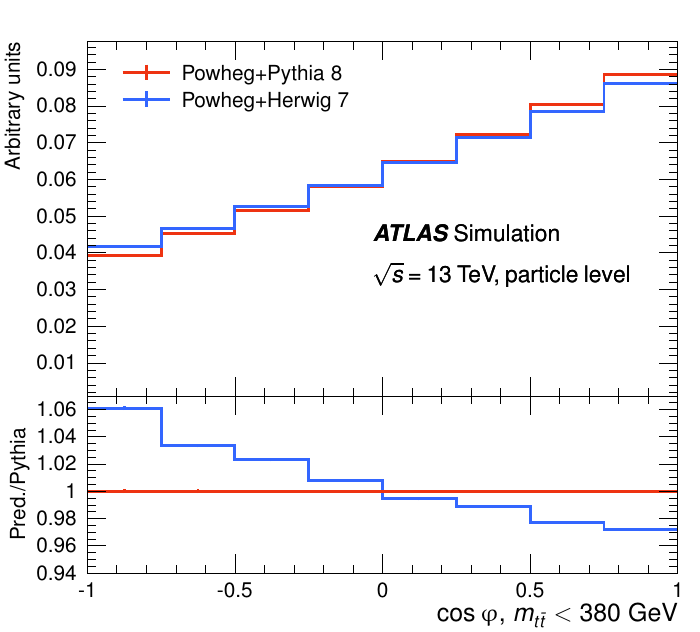}
\label{fig:PSHAD_Direct3a}
}
\subfigure[]{
\includegraphics[width=0.435\textwidth]{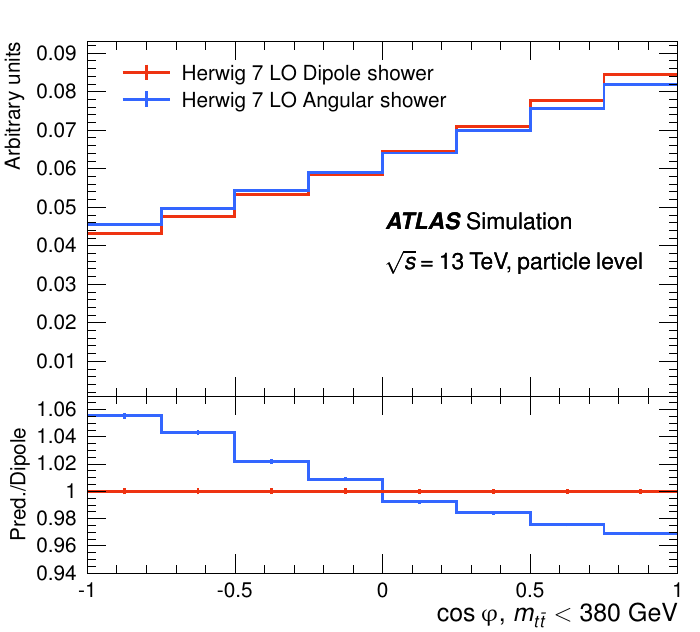}
\label{fig:PSHAD_Direct3b}
}
\caption{Comparison between \cosphi distributions in the signal region with $\mttbar < 380$~\GeV\ for different MC event generator setups at stable-particle level. Figure (a) compares events simulated with \POWHEGBOX which are interfaced with either \PYTHIA (red line, \pT-ordered dipole shower) or \HERWIG (blue line, angular-ordered shower) while figure (b) compares events simulated with \HERWIG using either a dipole-ordered shower (red line) or an angular-ordered shower (blue line).}
\label{fig:PSHAD_Direct3}
\end{figure}

\section*{Author contributions}
\vspace{-5pt}
All authors have contributed to the publication, being variously involved in the design and the construction of the detectors, in writing software, calibrating subsystems, operating the detectors and acquiring data, and finally analysing the processed data. The ATLAS Collaboration members discussed and approved the scientific results. The manuscript was prepared by a subgroup of authors appointed by the collaboration and subject to an internal collaboration-wide review process. All authors reviewed and approved the final version of the manuscript.
 
\section*{Competing financial interests}
\vspace{-5pt}
The authors declare no competing financial interests.
 
\section*{Data availability}
\vspace{-5pt}
Raw data were generated by the ATLAS experiment. Derived data supporting the findings of this study are available from the ATLAS collaboration upon request.
 
\section*{Code availability}
\vspace{-5pt}
The ATLAS data reduction software is available at the following link: \href{https://zenodo.org/record/4772550}{https://zenodo.org/record/4772550}.  Statistical modeling and analysis is based on the ROOT software and its embedded RooFit and RooStats modules, available at the following link:  \href{https://zenodo.org/record/3895852}{https://zenodo.org/record/4772550}.  Code to configure these statistical tools and to process their output is available upon request.


\printbibliography[heading=subbibliography]
\end{refsection}

\clearpage

\clearpage
\input{atlas_authlist}

\end{document}

%% file: atlas_authlist.tex
 
\begin{flushleft}
\hypersetup{urlcolor=black}
{\Large The ATLAS Collaboration}

\bigskip

\AtlasOrcid[0000-0002-6665-4934]{G.~Aad}$^\textrm{\scriptsize 102}$,
\AtlasOrcid[0000-0002-5888-2734]{B.~Abbott}$^\textrm{\scriptsize 120}$,
\AtlasOrcid[0000-0002-1002-1652]{K.~Abeling}$^\textrm{\scriptsize 55}$,
\AtlasOrcid[0000-0001-5763-2760]{N.J.~Abicht}$^\textrm{\scriptsize 49}$,
\AtlasOrcid[0000-0002-8496-9294]{S.H.~Abidi}$^\textrm{\scriptsize 29}$,
\AtlasOrcid[0000-0002-9987-2292]{A.~Aboulhorma}$^\textrm{\scriptsize 35e}$,
\AtlasOrcid[0000-0001-5329-6640]{H.~Abramowicz}$^\textrm{\scriptsize 151}$,
\AtlasOrcid[0000-0002-1599-2896]{H.~Abreu}$^\textrm{\scriptsize 150}$,
\AtlasOrcid[0000-0003-0403-3697]{Y.~Abulaiti}$^\textrm{\scriptsize 117}$,
\AtlasOrcid[0000-0002-8588-9157]{B.S.~Acharya}$^\textrm{\scriptsize 69a,69b,n}$
% ...
\bigskip
\\

% ...
$^{29}$Physics Department, Brookhaven National Laboratory, Upton NY; United States of America.\\
% ...
$^{35}$$^{(a)}$Facult\'e des Sciences Ain Chock, Universit\'e Hassan II de Casablanca;$^{(b)}$Facult\'{e} des Sciences, Universit\'{e} Ibn-Tofail, K\'{e}nitra;$^{(c)}$Facult\'e des Sciences Semlalia, Universit\'e Cadi Ayyad, LPHEA-Marrakech;$^{(d)}$LPMR, Facult\'e des Sciences, Universit\'e Mohamed Premier, Oujda;$^{(e)}$Facult\'e des sciences, Universit\'e Mohammed V, Rabat;$^{(f)}$Institute of Applied Physics, Mohammed VI Polytechnic University, Ben Guerir; Morocco.\\
% ...
$^{49}$Fakult\"{a}t Physik , Technische Universit{\"a}t Dortmund, Dortmund; Germany.\\
% ...
$^{55}$II. Physikalisches Institut, Georg-August-Universit\"{a}t G\"ottingen, G\"ottingen; Germany.\\
% ...
$^{69}$$^{(a)}$INFN Gruppo Collegato di Udine, Sezione di Trieste, Udine;$^{(b)}$ICTP, Trieste;$^{(c)}$Dipartimento Politecnico di Ingegneria e Architettura, Universit\`a di Udine, Udine; Italy.\\
% ...
$^{102}$CPPM, Aix-Marseille Universit\'e, CNRS/IN2P3, Marseille; France.\\
% ...
$^{117}$Department of Physics, New York University, New York NY; United States of America.\\
% ...
$^{120}$Homer L. Dodge Department of Physics and Astronomy, University of Oklahoma, Norman OK; United States of America.\\
% ...
$^{150}$Department of Physics, Technion, Israel Institute of Technology, Haifa; Israel.\\
$^{151}$Raymond and Beverly Sackler School of Physics and Astronomy, Tel Aviv University, Tel Aviv; Israel.\\
% ...
$^{a}$ Also Affiliated with an institute covered by a cooperation agreement with CERN.\\
$^{b}$ Also at An-Najah National University, Nablus; Palestine.\\
$^{c}$ Also at Borough of Manhattan Community College, City University of New York, New York NY; United States of America.\\
$^{d}$ Also at Center for High Energy Physics, Peking University; China.\\
$^{e}$ Also at Center for Interdisciplinary Research and Innovation (CIRI-AUTH), Thessaloniki; Greece.\\
$^{f}$ Also at Centro Studi e Ricerche Enrico Fermi; Italy.\\
% ...
$^{n}$ Also at Department of Physics, King's College London, London; United Kingdom.\\
% ...

\end{flushleft}
